\documentstyle[psfig,prb,multicol,aps]{revtex}

\begin{document}

\draft

\title{Long-range interaction and nonlinear localized
modes in photonic crystal waveguides}

\author{Serge F. Mingaleev $^{1,*}$, Yuri S. Kivshar $^1$,
and Rowland A. Sammut $^2$}

\address{$^1$ Optical Sciences Centre, Australian National 
University, Canberra ACT 0200,  Australia}

\address{$^2$ School of Mathematics and Statistics,
Australian Defence Force Academy, Canberra ACT 2600, Australia}

\maketitle

\begin{abstract}
We develop the theory of nonlinear localised modes ({\em intrinsic 
localised modes} or {\em discrete breathers}) in two-dimensional 
(2D) photonic crystal waveguides.  We consider different geometries 
of the waveguides created by an array of nonlinear dielectric rods 
in an otherwise perfect linear 2D photonic crystal, and demonstrate 
that the effective  interaction in such waveguides is {\em nonlocal}, 
being described by a new type of nonlinear lattice models with 
long-range coupling and nonlocal nonlinearity. We reveal the existence 
of different types of nonlinear guided modes  which are also localised 
in the waveguide direction, and describe their unique properties 
including bistability.
\end{abstract}

%

\begin{multicols}{2}
\narrowtext

\section{Introduction}

In physics, the idea of localisation is generally associated with 
disorder that breaks translational invariance. However, research in 
recent years has demonstrated that localisation can occur in the 
absence of any disorder and solely due to nonlinearity, in the 
form of {\em intrinsic localised modes}, also called {\em discrete 
breathers}.\cite{review} A rigorous proof of the existence of
time-periodic, spatially localised solutions describing such 
nonlinear modes has been presented for a broad class of
Hamiltonian coupled-oscillator nonlinear lattices,\cite{mak} but
approximate analytical solutions can also be found in many other
cases, demonstrating a generality of the concept of {\em nonlinear 
localised modes}.

Nonlinear localised modes can be easily identified in numerical 
molecular-dynamics simulations in many different physical models 
(see, e.g., Ref. \onlinecite{review} for a review), but only very 
recently the first experimental observations of spatially localised 
nonlinear modes have been reported in mixed-valence transition metal 
complexes,\cite{bishop} quasi-one-dimensional antiferromagnetic 
chains,\cite{sievers} and arrays of Josephson junctions.\cite{JJ} 
Importantly, 
very similar types of spatially localised nonlinear modes have been 
experimentally observed in {\em macroscopic} mechanical 
\cite{zolo} and guided-wave optical \cite{silb} systems.

Recent experimental observations of nonlinear localised modes, as
well as numerous theoretical results, indicate that both effects, i.e.
nonlinearity-induced localisation and spatially localised modes, can be
expected in physical systems of very different nature. From the
viewpoint of possible practical applications,  self-localised states in
optics  seem to be the most promising ones; they can lead to different 
types of nonlinear all-optical
switching devices where light manipulates and controls light itself, by
varying the input intensity. As a result, the study of nonlinear localised
modes in photonic structures is expected to bring  a variety of
realistic applications of intrinsic localised modes.

One of the promising fields where the concept of nonlinear localised modes
may find practical applications is in the physics of {\em photonic
crystals} [or photonic band gap (PBG) materials] --- 
periodic dielectric structures that produce
many of the same phenomena for photons as the crystalline atomic potential
does for electrons.\cite{pbg} Three-dimensional (3D) photonic crystals for
visible light have been successfully fabricated only within the past year
or two, and presently many research groups are working on creating tunable
band-gap switches and transistors operating entirely with light. The most
recent idea is to employ nonlinear properties of band-gap materials, thus
creating {\em nonlinear photonic crystals} that have 2D or 3D periodic
nonlinear susceptibility.\cite{berger,sukh}

Nonlinear photonic crystals or photonic crystals with embedded nonlinear
impurities create an ideal environment for the generation and observation
of nonlinear localised photonic modes. In particular, such modes can be
excited at nonlinear interfaces with quadratic nonlinearity,\cite{sukh2} 
or along dielectric waveguide structures possessing a nonlinear
Kerr-type response.\cite{mcgurn} In this paper, we analyse nonlinear
localised modes in 2D photonic crystal waveguides. We consider the
waveguides created by an array of dielectric rods in an otherwise perfect
2D photonic crystal. It is assumed that the dielectric constant of the
waveguide rods depends on the field intensity (due to the Kerr effect),
so that waveguides of different geometries can support a variety of
nonlinear guided modes. We demonstrate here that localisation can occur in
the propagation direction creating a 2D spatially localised mode 
(see Fig. \ref{fig:sol2d} below). As
follows from our results, the effective interaction in such nonlinear
waveguides is nonlocal, and the nonlinear localised modes are described by
a nontrivial generalisation of nonlinear lattice models with long-range
coupling and nonlocal nonlinearity.

\section{Model}

We consider a 2D photonic crystal created by a square lattice
of parallel, infinitely long dielectric columns (or rods) in air. 
The system is characterized by the dielectric constant 
$\epsilon(\bbox{x})=\epsilon(x_1, x_2)$, and it is assumed that 
the rods are parallel to the $x_3$ axis.  The evolution of the 
TM-polarised \cite{pbg} light [with the electric field having 
the structure $\bbox{E}=(0,0,E)$], propagating in 
the $(x_1, x_2)$-plane, is governed by the 
scalar wave equation
\begin{equation}
\nabla^2 E(\bbox{x}, t) - \frac{1}{c^2} \, \partial_t^2
\left[ \epsilon(\bbox{x}) E \right] = 0 \; ,
\label{sys:eq-E-t}
\end{equation}
where
$\nabla^2 \equiv \partial_{x_1}^2 + \partial_{x_2}^2$.
For monochromatic light, we consider the stationary solutions
\begin{displaymath}
E(\bbox{x}, t) = e^{-i \omega t} \, E(\bbox{x} \,|\, \omega) \; ,
\label{sys:E-t-omega}
\end{displaymath}
and the equation of motion (\ref{sys:eq-E-t}) reduces to the 
simple eigenvalue problem
\begin{equation}
\left[ \nabla^2  + \left( \frac{\omega}{c} \right)^2
\epsilon(\bbox{x}) \right]
E(\bbox{x} \,|\, \omega) = 0 \; .
\label{sys:eq-E-omega}
\end{equation}
This eigenvalue problem can be easily solved 
(e.g., by the plane waves method \cite{Maradudin:1993:PBGL}) in the 
case of a perfect photonic crystal, for which the dielectric constant 
$\epsilon(\bbox{x}) \equiv \epsilon_{pc}(\bbox{x})$ is a periodic 
function 
\begin{equation}
\epsilon_{pc}(\bbox{x}+\bbox{s}_{ij}) =
\epsilon_{pc}(\bbox{x}) \; , 
\label{sys:eps-pc}
\end{equation} 
where $i$ and $j$ are arbitrary integer, and 
\begin{equation}
\bbox{s}_{ij} = i \, \bbox{a}_1 + j \, \bbox{a}_2
\label{sys:s-ij}
\end{equation}
is a linear combination of the primitive lattice vectors
$\bbox{a}_1$ and $\bbox{a}_2$ of the 2D photonic crystal.

\begin{figure}
\centerline{\hbox{
\psfig{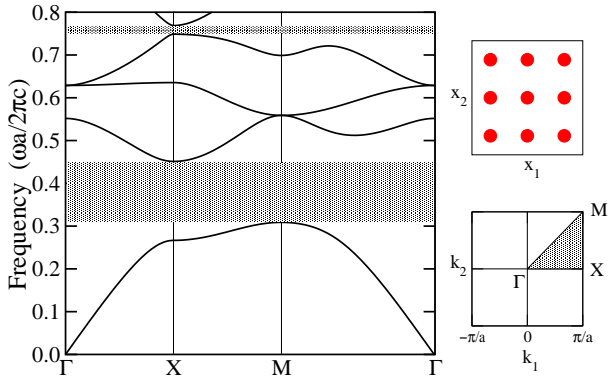}}}
\vspace{3mm}
\caption{The band-gap structure of the photonic crystal consisting 
of a square lattice of dielectric rods with $r_0=0.18a$ and 
$\epsilon_0=11.56$ (the band gaps are shaded grey). 
The top right inset shows a cross-sectional 
view of the 2D photonic crystal. 
The bottom right inset shows the corresponding Brillouin zone, 
with the irreducible zone shaded grey.}
\label{fig:band-r0.18}
\end{figure}

For definiteness, we consider the 2D photonic crystal
earlier analysed (in the linear limit) 
in Refs. \onlinecite{Mekis:1996:PRL,Mekis:1998:PRB}, 
i.e. we assume that the rods are identical and cylindrical, with 
radius $r_0=0.18a$ and dielectric constant 
$\epsilon_0=11.56$. The rods form a perfect square lattice with the
distance $a$ between two neighbouring rods, i.e.
$\bbox{a}_1=a \bbox{x}_1$ and $\bbox{a}_2=a \bbox{x}_2$.
The frequency band structure for this type of 2D photonic crystal, 
and for the selected polarisations of the electric field, is shown in 
Fig. \ref{fig:band-r0.18}.  As follows from the structure of the 
frequency spectrum, there 
exists a large (38\%) band gap that extends from the lower cut-off 
frequency, $\omega=0.302 \times 2\pi c/a$, to the upper band-gap 
frequency, $\omega=0.443 \times 2\pi c/a$. Since the characteristics 
of a PBG material remain unchanged under rescaling, we can assume 
that this gap is created either in the infra-red or visible 
regions of the spectrum. For example, if we choose the lattice 
constant to be $a =0.58 \, \mu$m, the wavelength corresponding to the 
mid-gap frequency  will be $1.55 \, \mu$m.

The TM-polarised light cannot propagate through the photonic crystal if its 
frequency falls inside the band gap. But one can excite guided modes 
inside the forbidden frequency gap by introducing some 
interfaces, waveguides, or defects. 
Here, we consider waveguides created by a row of identical defects 
with a Kerr-type nonlinear response.
These defect-induced waveguides possess translational symmetry,
and the corresponding guided modes can be characterized by the
reciprocal space wave vector $k$ directed along the waveguide.
Such a guided mode has a periodical profile inside the waveguide,
and it decays exponentially outside it.

{\em Linear photonic-crystal waveguides} created by removing a row of 
dielectric rods have been recently investigated numerically
\cite{Mekis:1996:PRL,Mekis:1998:PRB} 
and experimentally.\cite{Lin:1998:SCI}  
In particular, highly efficient transmission
of light, even in the case of a bent waveguide, has been demonstrated.

In the present paper, in contrast to Refs. 
\onlinecite{Mekis:1996:PRL,Mekis:1998:PRB,Lin:1998:SCI} 
where only linear waveguides were considered,  we study the 
properties of  {\em nonlinear waveguides} created by inserting an 
additional row of rods fabricated from a Kerr-type nonlinear 
material characterized by a third-order nonlinear susceptibility 
with the linear  dielectric constant $\epsilon_d$.  For definiteness, 
we assume that $\epsilon_d = \epsilon_0 = 11.56$. As we show below, 
changing the radius $r_d$ of these defect rods and their location
within the crystal, we can create waveguides 
with quite different properties.

\section{ Effective discrete equations}

Writing the dielectric constant $\epsilon(\bbox{x})$ as a sum of 
periodic and defect-induced terms, i.e. 
\begin{displaymath}
\epsilon(\bbox{x})=\epsilon_{pc}(\bbox{x})+\delta
\epsilon(\bbox{x} \,|\, E) \; ,
\label{sys:eps}
\end{displaymath}
we can present Eq. (\ref{sys:eq-E-omega}) as follows,
\begin{eqnarray}
\left[ \nabla^2 + \left( \frac{\omega}{c} \right)^2
\epsilon_{pc}(\bbox{x}) \right] &&
E(\bbox{x} \,|\, \omega) \nonumber \\
= - && \left( \frac{\omega}{c} \right)^2
\delta \epsilon(\bbox{x} \,|\, E) \, E(\bbox{x} \,|\, \omega) \; .
\label{sys:eq-E-omega-delta}
\end{eqnarray}
Equation (\ref{sys:eq-E-omega-delta}) can also be written in the 
integral form
\begin{equation}
E(\bbox{x} \,|\, \omega) = \left( \frac{\omega}{c} \right)^2
\!\!
\int d^2\bbox{y} \,\,\, G(\bbox{x}, \bbox{y} \,|\, \omega) \,
\delta \epsilon(\bbox{y} \,|\, E) \, E(\bbox{y} \,|\, \omega) \; ,
\label{sys:eq-green-int}
\end{equation}
where $G(\bbox{x}, \bbox{y} \,|\, \omega)$ is the Green function 
which is defined, in a standard way, as a  solution of the equation
\begin{displaymath}
\left[ \nabla^2 + \left( \frac{\omega}{c} \right)^2
\epsilon_{pc}(\bbox{x}) \right]
G(\bbox{x}, \bbox{y} \,|\, \omega) = - \delta(\bbox{x}-\bbox{y}) \; ,
\label{sys:eq-green-omega}
\end{displaymath}
with, accordingly to Eq. (\ref{sys:eps-pc}), periodic
coefficients. The properties of the Green 
function and the numerical methods for its calculation 
have been already described in the literature.
\cite{Maradudin:1993:PBGL,Ward:1998:PRB} 
Here, we notice that the Green function 
of a perfect 2D photonic crystal is {\em symmetric}, i.e.
\begin{displaymath}
G(\bbox{x}, \bbox{y} \,|\, \omega) =
G(\bbox{y}, \bbox{x} \,|\, \omega)
\label{sys:green-symm}
\end{displaymath}
and {\em periodic}, i.e.
\begin{displaymath}
G(\bbox{x} + \bbox{s}_{ij}, \bbox{y}  + \bbox{s}_{ij} \,|\, \omega) =
G(\bbox{x}, \bbox{y} \,|\, \omega) \; ,
\label{sys:green-period}
\end{displaymath}
where $\bbox{s}_{ij}$ is defined by Eq. (\ref{sys:s-ij}).

Let us consider a row of {\em nonlinear defect rods} embedded into 
the crystal along a selected direction. 
To describe such a row, we should define the rods 
positions along $\bbox{s}_{ij}$ with some specific values of 
$i$ and $j$. For example, let us first assume that the defect rods 
are located at the points
$\bbox{x}_m = \bbox{x}_0 + m \, \bbox{s}_{ij}$.
In this case, the correction to the dielectric constant is
\begin{eqnarray}
\delta \epsilon(\bbox{x}) = \left\{\epsilon_{d} +
|E(\bbox{x} \,|\, \omega)|^2\right\}
\sum_m \theta (\bbox{x}-\bbox{x}_m) \; ,
\label{sys:delta-eps}
\end{eqnarray}
where
\begin{displaymath}
\theta (\bbox{x}) = \left\{
\begin{array}{c}
1 \; , \quad \mbox{for} \quad |\bbox{x}| \leq r_d \; , \\
0 \; , \quad \mbox{for} \quad |\bbox{x}| > r_d \; .
\end{array}
\right.
\end{displaymath}
Assuming that the radius of the rods, 
$r_{d}$, is sufficiently small
(so that the electric field $E(\bbox{x} \,|\, \omega)$
is almost constant inside the defect rods), we substitute
Eq. (\ref{sys:delta-eps}) into Eq. (\ref{sys:eq-green-int}) and,
averaging over of the cross-section of the rods, derive an 
approximate {\em discrete nonlinear equation} for the electric field
\begin{eqnarray}
E_n = \sum_m J_{n-m}(\omega) (\epsilon_{d} + |E_m|^2)
E_m \; ,
\label{sys:eq-E-disc}
\end{eqnarray}
where
\begin{equation}
J_{n}(\omega) = \left( \frac{\omega}{c} \right)^2
\int\limits_{r_d} d^2 \bbox{y} \,\,\,
G(\bbox{x}_0, \bbox{x}_n + \bbox{y} \,|\, \omega ) \; .
\label{sys:Jn}
\end{equation}
This type of discrete nonlinear equation for photonic crystals 
has been earlier introduced by McGurn \cite{mcgurn},  for the 
special case of nonlinear impurities embedded in the linear rods. 
However, the analytical approach developed by McGurn for that model 
did not take into account the field distribution via the 
explicit dependence of the coupling coefficients  $J_n(\omega)$ and, 
as a result, the equation (\ref{sys:eq-E-disc}) was not solved
exactly. Moreover, the analysis of Ref. \onlinecite{mcgurn} 
was based on the nearest-neighbour 
approximation where the coupling coefficients are approximated as 
$J_n= J_0\delta_{n,0} + J_1 \delta_{n,\pm 1}$ with constant 
$J_0$ and $J_1$.

\begin{figure}
\centerline{\hbox{
\psfig{figure=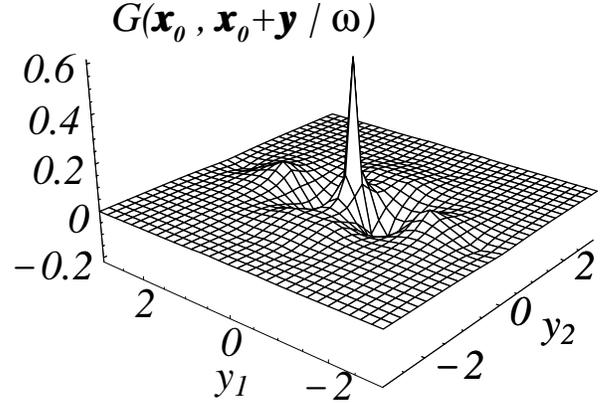,clip=,width=80mm,angle=0}}}
\caption{The Green function $G(\bbox{x}_0, \bbox{x}_0 + \bbox{y}
\,|\, \omega)$ for $\bbox{x}_0=\bbox{a}_1/2$ and
$\omega=0.33 \times 2\pi c/a$.}
\label{fig:green-r0.18}
\end{figure}

In a sharp contrast, in the present paper we provide a systematic 
numerical analysis of different types of 
nonlinear localised modes in the framework of a complete 
model. In particular, we reveal 
that the approximation of the nearest-neighbour interaction is very 
crude in many of the cases analysed.
Since the effective coupling coefficients are defined by the Green 
function, this can be seen directly from Fig. \ref{fig:green-r0.18}
that shows a typical spatial profile of the Green function which,
in general, characterises a long-range interaction, very typical 
for photonic crystal waveguides. As a consequence of that, the 
coupling coefficients $|J_n(\omega)|$ calculated 
from  Eq. (\ref{sys:Jn}) decrease exponentially with the site number 
$n$, and in the asymptotic region they can be presented as follows 
\begin{displaymath}
|J_n(\omega)| \approx \left\{
\begin{array}{lcc}
J_0(\omega) \; , & \mbox{for} & n=0 \; , \\
J_{*}(\omega) \, e^{-\alpha(\omega) |n|} \; ,
& \mbox{for} & |n| \geq 1 \; ,
\end{array}
\right.
\label{sys:Jn-exp}
\end{displaymath}
where the characteristic decay rate $\alpha(\omega)$ can be as small 
as $0.85$, depending on the values of $\omega$, $\bbox{x}_0$, 
$\bbox{s}_{ij}$, and $r_d$, and it can be even smaller for other 
types of photonic crystals.

This result allows us to draw an analogy with a class of the nonlinear 
Schr\"odinger (NLS) equations that describe nonlinear excitations 
in quasi-one-dimensional molecular chains with long-range (e.g. 
dipole-dipole) interaction between the particles and local on-site 
nonlinearities.\cite{Johansson:1998:PRE} For such systems, it was shown 
that the effect of nonlocal interparticle interaction introduces 
some new features in the properties of 
existence and stability of nonlinear 
localised modes. Moreover, for our model 
the coupling coefficients $J_n(\omega)$ can be either non-staggered 
and monotonically decaying, i.e. $J_n(\omega)=|J_n(\omega)|$,  or 
staggered and oscillating from site to site, i.e. 
$J_n(\omega)=(-1)^{n}|J_n(\omega)|$. We can therefore expect that 
effective nonlocality in both linear and nonlinear terms of 
Eq. (\ref{sys:eq-E-disc}) will bring a number of new features 
in the properties of nonlinear localised modes.

\section{Examples of nonlinear modes}

\begin{figure}
\centerline{\hbox{
\psfig{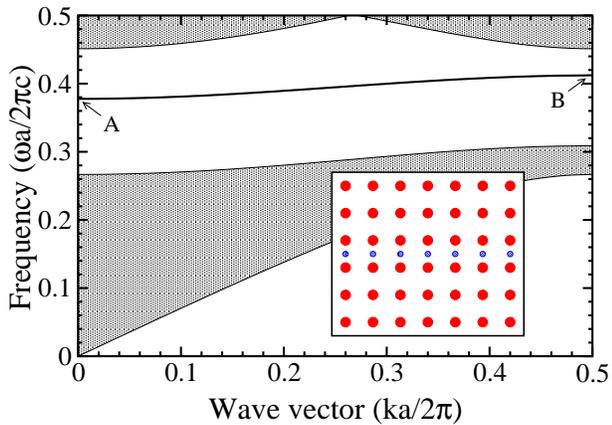}}}
\vspace{3mm}
\caption{Dispersion relation for the photonic crystal waveguide 
shown in the inset ($\epsilon_0=\epsilon_d=11.56$, $r_0=0.18a$,
$r_d=0.10a$). The grey areas are the projected band
structure of the perfect 2D photonic crystal. The frequencies at 
the indicated points are: $\omega_A=0.378 \times 2\pi c/a$ and 
$\omega_B=0.412  \times 2\pi c/a$.}
\label{fig:def-x2-0.10}
\end{figure}

As can be seen from the structure of the example Green function,
presented in Fig. \ref{fig:green-r0.18},  the case of monotonically
varying $J_n(\omega)$ can be obtained by locating the defect
rods at the points $\bbox{x}_0=\bbox{a}_1/2$,  along the straight
line in the $\bbox{s}_{01}$ direction. In this case, the frequency of 
a linear guided mode, that can be excited in such a waveguide, takes
the minimum value at $k=0$ (see Fig. \ref{fig:def-x2-0.10}), and the 
corresponding nonlinear mode is expected to be non-staggered.

\begin{figure}
\centerline{\hbox{
\psfig{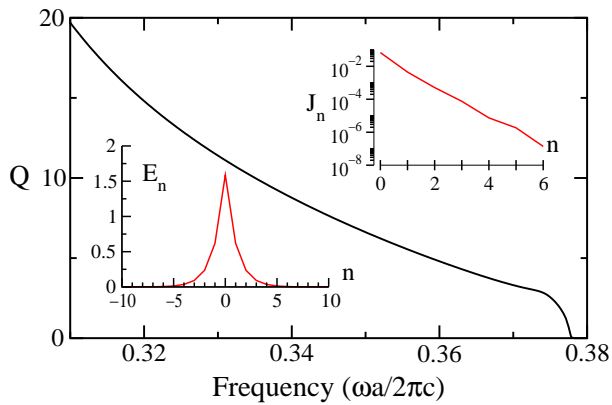}}}
\vspace{3mm}
\caption{Intensity $Q(\omega)$ of the nonlinear mode excited in 
the photonic crystal 
waveguide shown in Fig. \protect\ref{fig:def-x2-0.10}. The right 
inset gives the dependence $J_n(\omega)$ calculated at 
$\omega=0.37 \times 2\pi c/a$. The left inset presents the profile 
of the corresponding nonlinear localised mode.}
\label{fig:norm-x2-0.10}
\end{figure}

We have solved Eq. (\ref{sys:eq-E-disc}) numerically and found 
that nonlinearity can lead to the existence of a new type of 
guided modes which are localised in both directions, i.e. in 
the direction perpendicular to the waveguide, due to the guiding 
properties of a channel created by 
defect rods, and in the direction of the waveguide, 
due to the self-trapping effect. Such nonlinear modes exist with 
frequencies below the frequency of the linear guided mode of the 
waveguide, i.e. below the frequency $\omega_A$ in 
Fig. \ref{fig:def-x2-0.10}, and are indeed 
non-staggered, with the bell-shaped profile along the waveguide 
direction shown in the left inset of 
Fig. \ref{fig:norm-x2-0.10}.

\begin{figure}
\centerline{\hbox{
\psfig{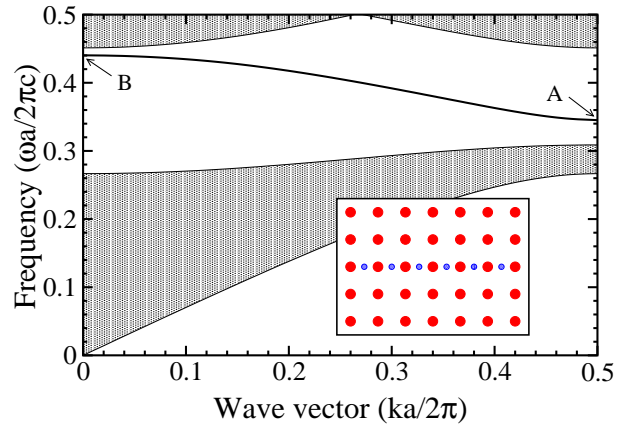}}}
\vspace{3mm}
\caption{Dispersion relation for the photonic crystal waveguide 
shown in the inset ($\epsilon_0=\epsilon_d=11.56$, $r_0=0.18a$,
$r_d=0.10a$). The grey areas are the projected band
structure of the perfect 2D photonic crystal. The frequencies 
at the indicated points are: $\omega_A=0.346 \times 2\pi c/a$ 
and $\omega_B=0.440 \times 2\pi c/a$.}
\label{fig:def-x1-0.10}
\end{figure}

\begin{figure}
\centerline{\hbox{
\psfig{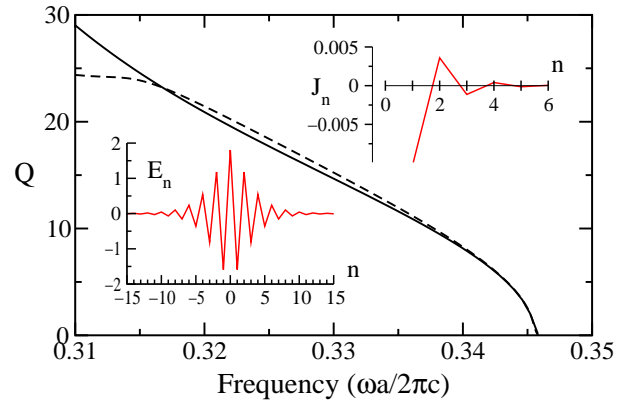}}}
\vspace{3mm}
\caption{Intensity $Q(\omega)$ of the nonlinear mode excited in 
the photonic crystal 
waveguide shown in Fig. \protect\ref{fig:def-x1-0.10}. The solid 
curve corresponds to the case of nonlinear rods in a linear 
photonic crystal, whereas the dashed curve is the same dependence 
for the case of  a nonlinear photonic crystal. The right inset 
shows the behaviour of the coupling coefficients $J_n(\omega)$ for 
$n \geq 1$ ($J_0=0.045$) at $\omega=0.33 \times 2\pi c/a$.
The left inset shows the profile of the corresponding nonlinear mode.}
\label{fig:norm-x1-0.10}
\end{figure}

The 2D nonlinear modes localised in both dimensions can be 
characterized by the mode intensity which we define, by analogy 
with the NLS equation, as
\begin{equation}
Q = \sum_n |E_n|^2.
\label{sys:norm}
\end{equation}
This intensity is closely related to the energy of
the electric field in the 2D photonic crystal accumulated in the nonlinear 
mode.  In Fig. \ref{fig:norm-x2-0.10} we plot the dependence of 
$Q$ on frequency, for the 
waveguide geometry shown in Fig. \ref{fig:def-x2-0.10}.

\begin{figure}
\centerline{\hbox{
\psfig{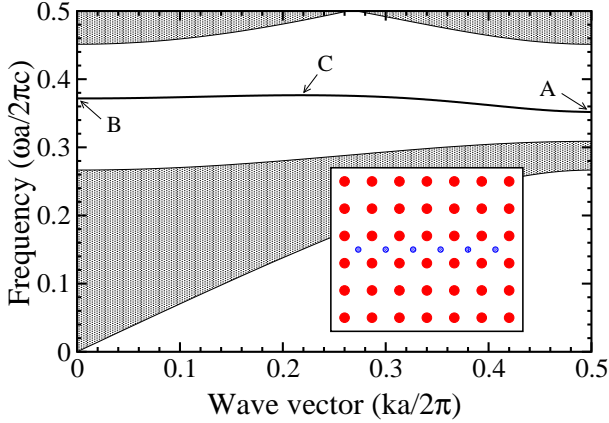}}}
\vspace{3mm}
\caption{Dispersion relation for the photonic crystal waveguide 
shown in the inset ($\epsilon_0=\epsilon_d=11.56$, $r_0=0.18a$,
$r_d=0.10a$). The grey areas are the projected band
structure of the perfect crystal. The frequencies at the points
indicated are: 
$\omega_A=0.352 \times 2\pi c/a$, $\omega_B=0.371 \times 
2\pi c/a$, and $\omega_C=0.376 \times 2\pi c/a$ (at $k=0.217 
\times 2\pi/a$).}
\label{fig:def-x12-0.10}
\end{figure}

\begin{figure}
\centerline{\hbox{
\psfig{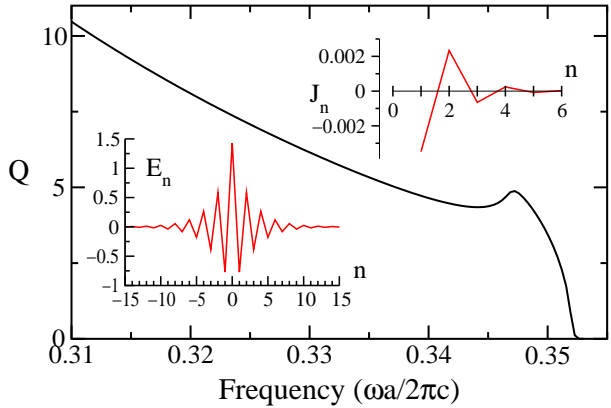}}}
\vspace{3mm}
\caption{Intensity $Q(\omega)$ of the nonlinear mode excited in 
the photonic crystal 
waveguide shown in Fig. \protect\ref{fig:def-x12-0.10}. 
The right inset shows the behaviour of the coupling coefficients 
$J_n(\omega)$ for
$n \geq 1$ ($J_0=0.068$) at  $\omega=0.345 \times 2\pi c/a$.
The left inset shows the profile of the corresponding nonlinear mode.}
\label{fig:norm-x12-0.10}
\end{figure}

As can be seen from the example of the Green 
function shown in Fig. \ref{fig:green-r0.18}, the case of staggered 
coupling coefficients $J_n(\omega)$ can be obtained by locating the 
defect rods at the points $\bbox{x}_0=\bbox{a}_1/2$, along the 
straight line in the $\bbox{s}_{10}$ direction. In this case, 
the frequency dependence of the linear guided mode of the waveguide 
takes the minimum at $k=\pi/a$ (see Fig. \ref{fig:def-x1-0.10}). 
Accordingly, the nonlinear guided mode localised along  the 
direction of the waveguide is expected to exist with the frequency 
below the lowest frequency $\omega_A$ of the linear guided mode, 
with a staggered profile.  The longitudinal profile of such a 2D 
nonlinear localised mode is shown in the left inset in 
Fig. \ref{fig:norm-x1-0.10}, together with the dependence of the 
mode intensity $Q$ on the frequency (solid curve), which in 
this case is again monotonic.

\begin{figure}
\centerline{\hbox{
\psfig{figure=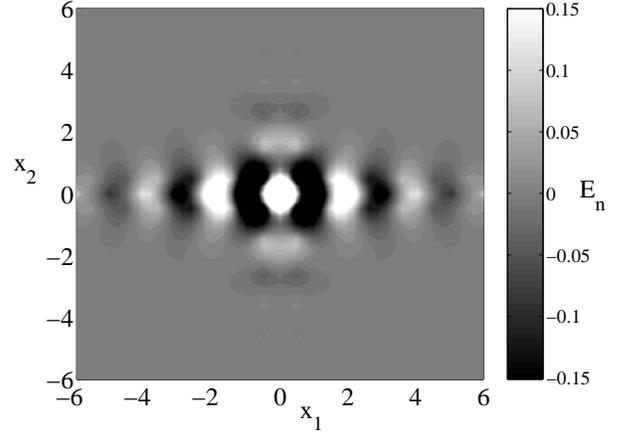,clip=,width=80mm,angle=0}}}
\vspace{3mm}
\caption{Contour plot of the the nonlinear localised mode that
corresponds to the longitudinal cross-section shown in 
the left inset of Fig. \protect\ref{fig:norm-x12-0.10}.}
\label{fig:sol2d}
\end{figure}

The results presented above are obtained for linear photonic 
crystals with nonlinear waveguides created by a row of defect 
rods. However, we have carried out the same analysis for the general 
case of {\em a nonlinear photonic crystal} that is created by 
rods of different size but made of the same nonlinear material. 
Importantly, we have found very small difference in all the results 
for relatively weak nonlinearities. In particular, 
for the photonic crystal waveguide shown in Fig. 
\ref{fig:def-x1-0.10}, the results for linear 
and nonlinear photonic crystals are very close. Indeed, for the mode 
intensity $Q$ the results corresponding to a nonlinear photonic crystal 
are shown in Fig. \ref{fig:norm-x1-0.10} by a dashed curve, and 
for $Q<20$ this curve almost coincides with the solid curve
corresponding to the case of a nonlinear waveguide embedded into 
a linear photonic crystal.

Let us now consider the waveguide created by a row of defects 
which are located at the points
$\bbox{x}_0=(\bbox{a}_1+\bbox{a}_2)/2$, 
along a straight line in either the $\bbox{s}_{10}$ or 
$\bbox{s}_{01}$ directions. The results for this case are 
presented in Figs.
\ref{fig:def-x12-0.10}--\ref{fig:sol2d}. 
The coupling coefficients $|J_n|$ are described by a slowly decaying  
function of the site number $n$, so that the effective interaction 
decays on scales  much larger than those in the cases considered 
previously.  Similar to the NLS models with long-range
dispersive interactions \cite{Johansson:1998:PRE,Gaididei:1997:PRE}, 
for this type of nonlinear photonic crystal waveguide we find a 
non-monotonic behaviour of the mode intensity $Q(\omega)$ and, as a 
result, multi-valued dependence of the invariant $Q(\omega)$ for 
$\omega<0.347 \times 2\pi c/a$.  
Similar to the results earlier obtained for the nonlocal 
NLS models \cite{Johansson:1998:PRE}, we can expect here 
that nonlinear localised modes corresponding, in our notations, 
to the positive slope of the derivative 
$dQ/d\omega$ are unstable and will eventually decay or transform 
into modes of higher or lower frequency. Such a phenomenon is 
known as {\em bistability}, and in this problem it
occurs as  a direct manifestation of the nonlocality of the 
effective (linear and nonlinear) interaction between the defect 
rod sites.

\section{Conclusions}

Exploration of nonlinear properties of PBG materials
is a new direction of research, and it may open up a new class of
applications of photonic crystals for all-optical signal processing 
and switching,  allowing an effective way to create tunable band-gap 
structures operating entirely with light. Nonlinear photonic crystals,  
and nonlinear waveguides embedded into photonic structures with 
periodically modulated dielectric constant,  create an ideal 
environment for the generation and observation of nonlinear 
localised modes.

In the present paper, we have developed a consistent theory of 
nonlinear localised modes which can be excited in photonic crystal 
waveguides of different geometry. For several geometries of 2D 
waveguides, we have demonstrated that such modes
are described by a new type of nonlinear lattice models that include 
long-range interaction and effectively nonlocal nonlinear response. 
It is expected that the general features of nonlinear guided modes 
described here will be preserved in other types of photonic crystal 
waveguides. Our approach and results can also be useful to develop 
the theory of nonlinear two-frequency parametric localised modes 
in the recently fabricated 2D photonic crystals
with the second-order nonlinear susceptibility \cite{neal}.  
Additionally, similar types of nonlinear localised modes are 
expected in photonic crystal fibers \cite{russell} consisting of a 
periodic air-hole lattice that runs along the length of the fiber, 
provided the fiber core is made of a highly nonlinear material 
(see, e.g., Ref. \onlinecite{egg}).

\section*{Acknowledgments}

Yuri Kivshar is thankful to Costas Soukoulis for useful discussions 
and suggestions at the initial stage of this project.  The work 
has been partially supported by the Large Grant Scheme 
of the Australian Research Council, the Australian Photonics 
Cooperative Research Centre, and the Planning and Performance 
Foundation grant of the Institute of Advanced Studies.


\end{multicols}
\end{document}